\documentclass[aps,prl,twocolumn,showpacs,amsmath,amssymb,amsthm,superscriptaddress,groupedaddress]{revtex4}
\usepackage{graphics}
\usepackage{epsfig}
\usepackage[colorlinks=true,citecolor=blue,linkcolor=red,linktocpage=true,pagebackref=false]{hyperref}
\usepackage{soul} 
\usepackage[usenames, dvipsnames]{color}
\usepackage{braket}

\begin{document}
\title{
Half-levitons -- zero-energy excitations of a driven Fermi sea
}
\author{Michael Moskalets}
\email{michael.moskalets@gmail.com}
\affiliation{Department of Metal and Semiconductor Physics, NTU ``Kharkiv Polytechnic Institute", 61002 Kharkiv, Ukraine}

\date\today
\begin{abstract}
A voltage pulse of a Lorentzian shape carrying a half of the flux quantum  excites out of a zero-temperature Fermi sea an electron in a mixed state, which looks like a quasi-particle with an effectively fractional charge $e/2$. A prominent feature of such an excitation is a narrow peak in the energy distribution function laying exactly at the Fermi energy $ \mu$. Another spectacular feature is that the distribution function has symmetric tails as above as below $ \mu$, which results in a zero energy of an excitation. This sounds improbable since at zero temperature all available states below $ \mu$ are fully occupied. The resolution is lying in the fact that such a voltage pulse excites also electron-hole pairs which free some space below $ \mu$ and thus  allow  a zero-energy quasi-particle to exist. I discuss also how to address separately electron-hole pairs and a fractionally charged zero-energy excitation in experiment. 
\end{abstract}
\maketitle

{\it Introduction.--}
Recent realization of a triggered single-electron source \cite{Blumenthal:2007ho,Feve:2007jx,Fujiwara:2008gt,Kaestner:2008hp,Roche:2013jw,Connolly:2013hx,Dubois:2013ul,Rossi:2014kp,dHollosy:2015ez,vanZanten:2016vw} opens a new era for a coherent electronics \cite{Webb:1985tu,Chandrasekhar:1985up,Schuster:1997vv,Henny:1999tb,Oliver:1999ws,Ji:2003ck,Neder:2007jl,Yamamoto:2012bp} by allowing it to go quantum much like a quantum optics. 
The analogues of the famous quantum optics effects were successfully demonstrated with single electrons in solid state circuits such as partitioning of electrons \cite{Bocquillon:2012if,Dubois:2013ul,Fletcher:2013kt,Ubbelohde:2014eq} in Hanbury-Brown and Twiss geometry and quantum-statistical repulsion of electrons \cite{Bocquillon:2013dp,Dubois:2013ul} in Hong-Ou-Mandel geometry. 
Tomography of a single-electron state \cite{Jullien:2014ii} and a preparation of few-electron Fock states \cite{Fletcher:2013kt,Waldie:2015hy,Glattli:2016tr} are already reported. 

An essential difference from quantum optics is that single electrons are injected into an electron wave-guide with another electrons such as, for instance, a quantum Hall edge channel \cite{Klitzing:1980kw,Halperin:1982ej,Buttiker:1988fn}.
During such an injection the source can excite an electron system and the resulting   excitations can mask injected electrons. 
However if the protocol of injection is properly chosen \cite{Mahe:2010cp} no spurious excitations appear. 
This was clearly demonstrated theoretically \cite{Levitov:1996ie,Ivanov:1997kf,Keeling:2006hq}  and experimentally \cite{Dubois:2013ul} in the case where single electrons are excited by applying a voltage pulse $V(t)$ across a ballistic conductor. 
It was shown that a voltage pulse of a Lorentzian shape with quantized Faraday flux, $\varphi \equiv (e/ \hbar) \int dt V(t) = 2 \pi n$ (where $e$ is the electron charge, $ \hbar$ is  Plank's constant, $n$ is an integer), excites only $n$ electrons (or holes, if $n<0$) with no accompanying electron-hole pairs. 
These excitations were named levitons.\cite{Dubois:2013ul}
While if the flux is not quantized, $ \varphi \ne 2 \pi n$, then what is excited is rather a messy state with a divergent number of quasi-particles, both electrons and holes.  

\begin{figure}[t]
\centerline{
\includegraphics[width=80mm]{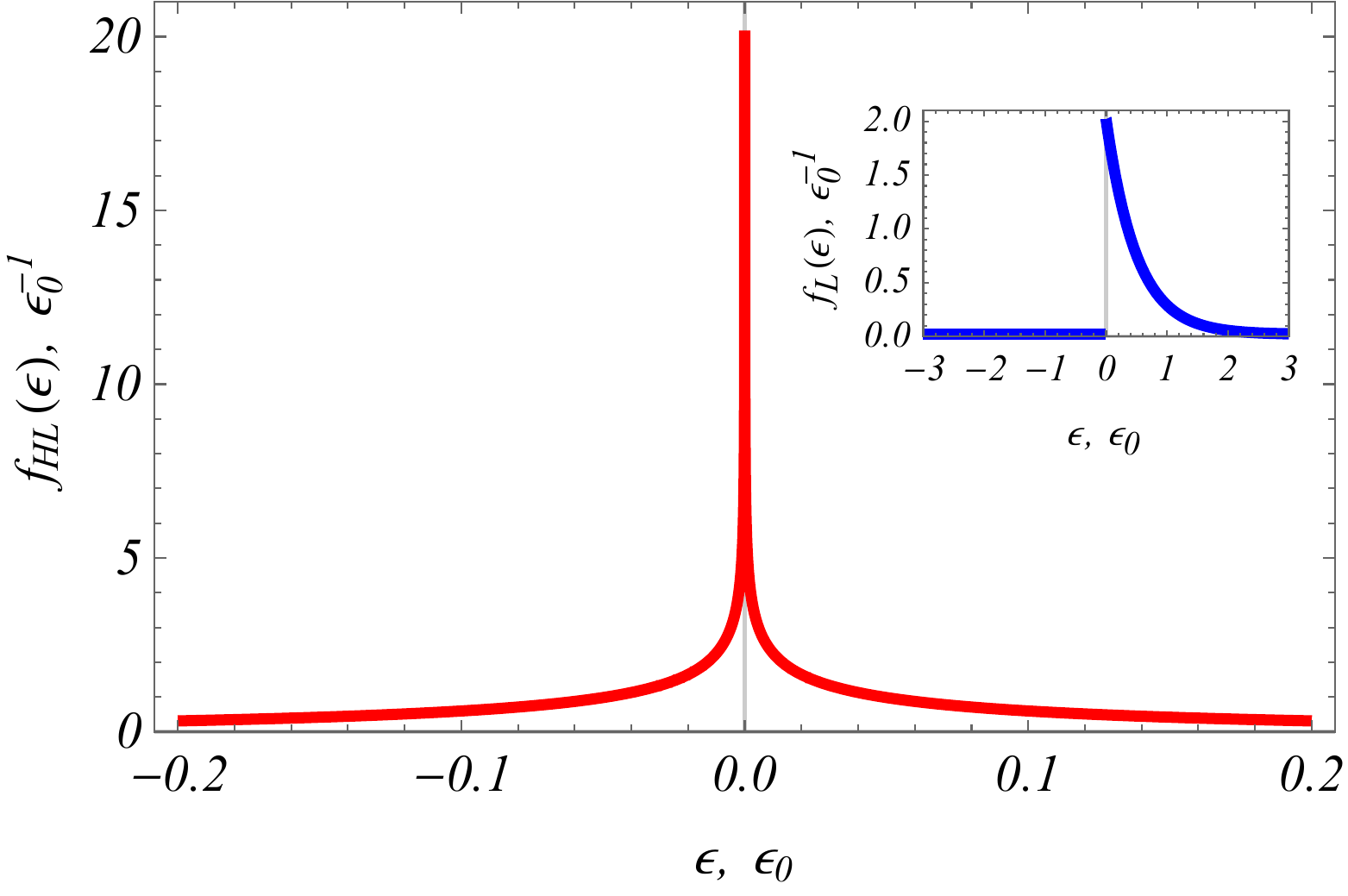}
}
\caption{(Color online) 
{\bf Main panel:} Energy distribution function $f_{HL}( \epsilon)$ of a half-leviton excited out of a zero-temperature Fermi sea with the help of a Lorentzian voltage pulse $V(t)$ carrying a half of the flux quantum, $ (e/ \hbar) \int dt V(t) = \pi$. The energy $ \epsilon = E - \mu $ is counted from the Fermi energy $ \mu$  and is normalized to $ \epsilon_{0} = \hbar/  \Gamma _{\tau}$ with $\Gamma _{\tau}$ being the half-width of a voltage pulse.  The peak at a zero energy is $f_{HL}( \epsilon\to 0) \approx \frac{2 }{ \pi^{2} \epsilon_{0} } \ln^{2}\left( \frac{ \epsilon_{0} }{ \left | \epsilon \right | } \right)$. {\bf Inset:} Energy distribution function of a leviton, a particle  with an integer charge $e$ excited by a voltage pulse $2v(t)$: $f_{L}( \epsilon>0) = (2/ \epsilon_{0}) \exp(- 2 \epsilon/ \epsilon_{0})$  and $f_{L}( \epsilon<0) =0$.\cite{Keeling:2006hq}}
\label{fig1}
\end{figure}

Here I show, however, that the flux $ \varphi = \pi $ is especial. 
The Fermi sea excited by a Lorentzian voltage pulse with a half-integer flux hosts an exotic single-particle excitation, which cannot exist in equilibrium, see Fig.~\ref{fig1}. 
Such an excitation has an effective charge $e/2$,  hence I name it {\it a half-leviton} (HL). 
Importantly, an electron-hole state (which is also excited because the flux is not quantized) is indispensable for existence of HLs. 
This is so since the state of a half-leviton is a superposition of states with energies laying from both sides of the Fermi energy $ \mu$, below and above it.  
At zero temperature all the state below the Fermi energy are fully occupied and only excited holes (belonging to electron-hole pairs) free some states below $ \mu$ and allow a half-leviton to be formed.

I stress that half-levitons are different from fractionally-charged clean pulses (FCCPs) in a Luttinger liquid, which were discussed in Refs.~\onlinecite{Keeling:2006hq,Jonckheere:2005kj}: 
For the existence of FCCPs an electron-electron interaction is crucial, while HLs can be excited in a non-interacting electron system; 
FCCPs can be excited alone, while HLs require necessarily  accompanying electron-hole pairs; 
FCCPs are fractional charge quasi-particles in a pure state, while HLs are rather quasi-particles with an integer charge $e$ being in a mixed state such that an effective charge is $e/2$; 
FCCPs carry a positive energy (counted from the Fermi energy), while HLs have a zero energy. 

\begin{figure}[b]
\centerline{
\includegraphics[width=80mm]{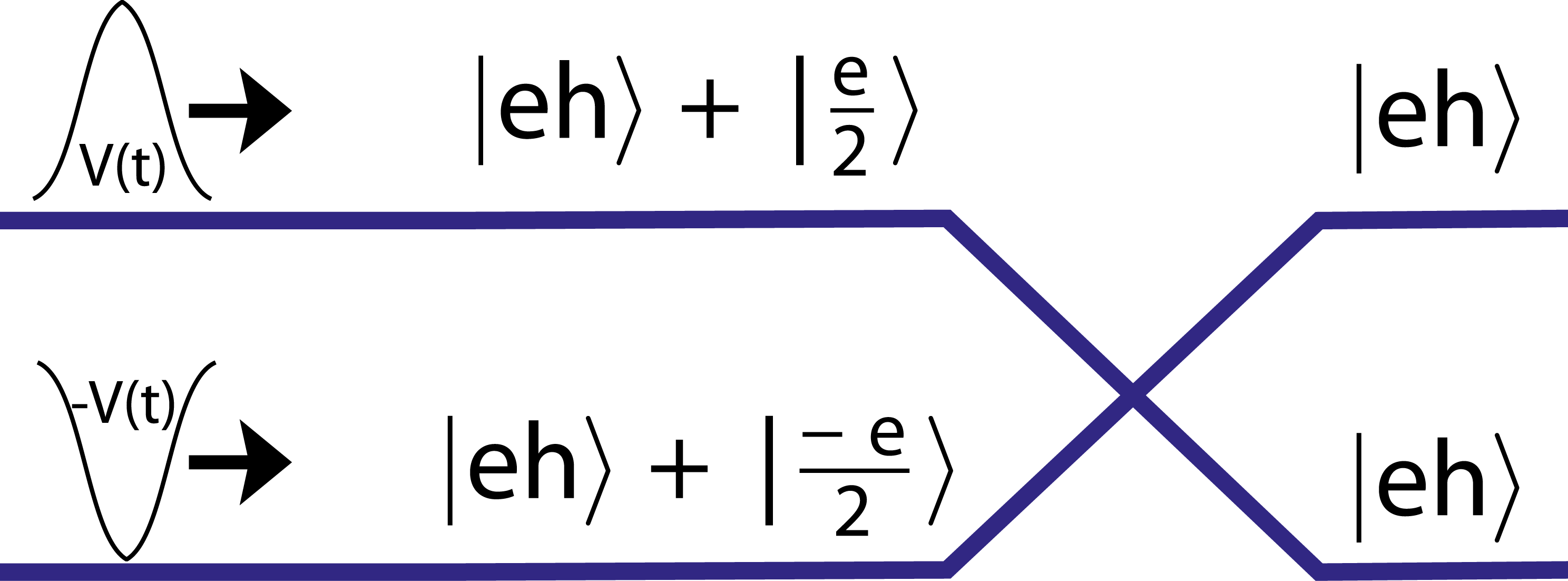}
}
\caption{(Color online) A sketch of an electronic wave splitter with colliding states excited by the Lorentzian voltage pulses of an opposite sign, $V(t)$ and $-V(t)$, carrying a half of the flux quantum each. One state is composed of electron-hole pairs, $\ket{eh}$, and  a half-leviton, $\ket{\frac{e }{2 }}$, and the other one is composed of electron-hole pairs and an anti-half-leviton, $\ket{\frac{-e }{2 }}$. The state projected onto one of the outputs of a symmetric wave splitter  contains only electron-hole pairs. 
}
\label{fig2}
\end{figure}

The last circumstance allows HL to annihilate (without breaking a phase coherence) its anti-particle, which is excited by a voltage pulse carrying a flux of an opposite sign, see Fig.~\ref{fig2}. 
Such a coherent annihilation on a wave splitter is impossible with ordinary quasi-particles, electrons and holes, whose energies lie above and below the Fermi energy, respectively. 
Therefore, they can be annihilated only as a result of inelastic processes, which generally break phase coherence. 
The elastic collisions of ordinary single electrons and holes do not lead to annihilation \cite{Juergens:2011gu,Jonckheere:2012cu,Hofer:2014jb,Dasenbrook:2015bm} unless in specific setups. 
For instance, where an electron emitted by one source is passed by and reabsorbed by the another source attempting to emit a hole.\cite{Splettstoesser:2008gc} 
Another example is a setup where energies of  electrons and holes are aligned but the success rate of annihilation is small.\cite{Beenakker:2004bv} 
A possibility for a perfect coherent annihilation of particles on a wave splitter predicted here opens a route for entangling Fock states with different number of fermions in solid-state quantum circuits.

{\it Half-leviton.--}
To characterize quasi-particles arising in a one-dimensional chiral or ballistic  system of non-interacting spinless electrons under the action of a dynamic excitation, an electron source, we introduce the excess first-order correlation function \cite{Grenier:2011js,Haack:2013ch}. 
This function is defined as the difference of electronic correlation functions with the source on and off,  
$G^{(1)}\left( 1; 2 \right) = \left\langle \hat \Psi^{\dag} \left( 1 \right) \hat \Psi \left( 2 \right) \right\rangle_{on} - \left\langle \hat \Psi^{\dag} \left( 1 \right) \hat \Psi \left( 2 \right) \right\rangle_{off}$. 
Here $\Psi \left( j \right)$ is an electron field operator calculated at point $x_{j}$ and time $t_{j}$ behind the source. 
The quantum statistical average $\left\langle \dots \right\rangle$ is taken over the equilibrium state of an electron system incoming to the place where the source is located. 
Incoming electrons are described by the Fermi distribution function with the chemical potential $ \mu$ and temperature $ \theta$.  
We will utilize the wide band approximation, when all the relevant energy scales are small compared to $ \mu$ and the spectrum of electrons of the Fermi sea can be linearized around the Fermi energy. 
In such a case the excess correlation function depends on a reduced time $t_{j} \equiv t_{j} - x_{j}/v_{ \mu}$ (with $v_{ \mu}$ the Fermi velocity) rather than on space and time coordinates separately.  

If quasi-particles are excited by a time-dependent voltage $V(t)$, then at zero temperature the excess correlation function is, \cite{Moskalets:2016dx}

\begin{eqnarray}
G ^{(1)}(t_{1}; t_{2}) =  \frac{e^{i \left( t_{1} - t_{2} \right) \frac{ \mu }{ \hbar } } }{  v_{ \mu} } \frac{  e^{ i \frac{e }{ \hbar} \int _{ t_{2} }^{t_{1} } dt^{\prime} V(t^{\prime}) } - 1 }{2 \pi i \left(  t_{1} - t_{2}  \right) } .
\label{G}
\end{eqnarray}

Here we are interested in a Lorentzian voltage pulse of width $  2\Gamma _{\tau}$, $eV(t) =  n^{\star}  2 \hbar \Gamma _{ \tau} / \left( t^2 + \Gamma _{ \tau}^2 \right)$, which caries a flux $ \varphi = 2 \pi  n^{\star}$. 
For $  n^{\star}= 0.5$ we get,

\begin{eqnarray}
G_{0.5} ^{(1)}(t_{1}; t_{2}) &=&  \frac{e^{i \left( t_{1} - t_{2} \right) \frac{ \mu }{ \hbar } } }{  v_{ \mu} } \left\{ g_{HL}\left( t_{1}; t_{2} \right)  + g_{eh}^{(1)} \left( t_{1}; t_{2} \right) \right\} ,
\nonumber \\ 
g_{HL}\left( t_{1}; t_{2} \right) &=& \frac{  \Gamma _{\tau} }{2 \pi\sqrt{ t_{1}^{2} +  \Gamma _{\tau}^{2} } \sqrt{t_{2}^{2} +  \Gamma _{\tau}^{2} } } ,
\label{HL} \\
g_{eh}^{(1)} \left( t_{1}; t_{2} \right) &=& \frac{ \frac{t_{1}t_{2} +  \Gamma _{\tau}^{2} }{\sqrt{ t_{1}^{2} +  \Gamma _{\tau}^{2} } \sqrt{t_{2}^{2} +  \Gamma _{\tau}^{2} } } -1 }{ 2 \pi i \left(  t_{1} - t_{2}  \right)} .
\nonumber 
\end{eqnarray}
\ \\ \noindent
The first term, $g_{HL}$, is factorized into the product of two terms dependent on a single time each. 
It describes a single-particle excitation since all corresponding higher-order correlation functions are identically zero.\cite{Moskalets:2015vr}
I call it a half-leviton because it is excited by a half voltage pulse, which excites a leviton \cite{Dubois:2013ul}, and mark corresponding quantities by a subscript $HL$. 
This excitation carries a charge $q^{\star}_{HL} = e \int dt g_{HL}\left( t; t \right) =  e/2$. 
To understand why a charge is fractional one needs to note that the state of HL  is a mixed state. 
This follows from the fact that the purity coefficient \cite{Moskalets:2016dx} calculated for $g_{HL}$ is less then one: $\mathrm{P}_{HL} = \int dt g_{HL}\left( t_{1}; t \right)g_{HL}\left( t; t_{2} \right)/ g_{HL}\left( t_{1}; t_{2} \right) = 0.5\,$. 
Since $q^{\star}_{HL} = e \mathrm{P}_{HL}$ one can say that the state in question corresponds to a single-particle with an integer charge $e$ appearing with probability $\mathrm{P}_{HL}$ and a vacuum state appearing with probability $1-\mathrm{P}_{HL}$. 
Threfore, $q^{\star}_{HL}$ is an effective charge. 

Using the purity coefficient we can write, $g_{HL}\left( t_{1}; t_{2} \right) = \mathrm{P}_{HL} \Phi^{*}_{HL}(t_{1}) \Phi_{HL}(t_{2})$, and find that a corresponding single-particle wave function $ \Phi_{HL}(t)$ can be chosen real-valued,

\begin{eqnarray}
\Phi_{HL}(t) = \sqrt{ \frac{  \Gamma _{\tau} }{ \pi }}  \frac{1 }{\sqrt{ t^{2} +  \Gamma _{\tau}^{2} } } .
\label{Phi}
\end{eqnarray}
\ \\ \noindent 
and normalized to one, $\int dt \left | \Phi_{HL}(t) \right |^{2} = 1$. 
Note that this wave function is symmetric in time, $\Phi_{HL}(t) = \Phi_{HL}(-t)$. 
 
The second term in Eq.~(\ref{HL}) describes electron-hole excitations (hence a subscript $eh$), which do not carry any charge, $I_{eh}(t) \equiv e g_{eh}^{(1)}(t;t) = 0$. 
Their presence can be verified via the shot noise measurement \cite{Bocquillon:2012if,Dubois:2013ul} or with the help of an interference current \cite{Gaury:2014jz,Moskalets:2014cg}).  
Electron-hole pairs do carry energy injected by a voltage pulse into an electron system. 
In contrast, HL does not carry any energy. 
To show this let us go over from time domain to energy domain and introduce the energy distribution function for excited particles, see, e.g. Ref.~\onlinecite{Moskalets:2014ea}: 

\begin{eqnarray}
f( \epsilon) &=& \frac{v_{ \mu} }{ h } \iint d t_{1} dt_{2} e^{- i( \mu + \epsilon) \frac{ t_{1} - t_{2} }{ \hbar }}  G^{(1)}\left( t_{1}; t_{2} \right),
\label{f}
\end{eqnarray}
\ \\ \noindent
where $ \epsilon$ is an energy counted from the Fermi energy. 
The function $f( \epsilon)$ is a probability density to find an excited particle with energy $ \epsilon$. 
Using a correlation function given in Eq.~(\ref{HL}) we find,

\begin{eqnarray}
f( \epsilon) &=& f_{HL}( \epsilon) + f_{eh}( \epsilon) ,
\nonumber \\
f_{HL}( \epsilon) &=& \frac{ \mathrm{P}_{HL} }{ h } \left | \int dt  \cos\left( \epsilon t /  \hbar  \right) \Phi_{HL}(t) \right |^{2} ,
\label{fHL} \\
f_{eh}( \epsilon) &=&  \frac{1 }{ h } \iint d t_{1} dt_{2} \sin\left( \epsilon [t_{2} - t_{1}]/ \hbar \right) i g_{eh}^{(1)}(t_{1}; t_{2}) .
\nonumber 
\end{eqnarray}
\ \\ \noindent
In the last equation I used $g_{eh}^{(1)}( - t_{1}; - t_{2} ) = g_{eh}^{(1)}( t_{1}; t_{2} )$.

The distribution function is normalized such that $ \int d \epsilon f( \epsilon) = \mathrm{P}_{HL}$. 
Electron-hole pairs do not contribute to this equation. 
The reason is the following. 
By virtue of definition, electron and hole contributions to the excess correlation function and, correspondingly, to the distribution function $f_{eh}( \epsilon)$ have opposite signs.  
As a result $\int d \epsilon f_{eh}( \epsilon) = 0$. 
To get separately the number of either electrons or holes we have to integrate $f_{eh}( \epsilon)$ over either positive or negative energies only.

The distribution function of a half-leviton is shown in Fig.~\ref{fig1}. 
This function is even in energy, $f_{HL}( \epsilon) = f_{HL}( - \epsilon)$ and, therefore, it does not contribute to the energy of excitations $ \left\langle \epsilon \right\rangle =  \left\langle \epsilon \right\rangle_{HL} +  \left\langle \epsilon \right\rangle_{eh}$,
\begin{eqnarray}
 \left\langle \epsilon \right\rangle_{HL} = \int d \epsilon f_{HL}( \epsilon) \epsilon = 0 .
\label{eHL}
\end{eqnarray}
This is why I call a half-leviton {\it a zero-energy excitation}. 
In contrast, a true leviton, excited by a voltage pulse with $  n^{\star} = 1$, has a non-zero energy, $ \left\langle \epsilon \right\rangle_{L} =  \int d \epsilon f_{L}( \epsilon) \epsilon=  \hbar/  (2 \Gamma _{\tau})$,\cite{Keeling:2006hq} see the inset to Fig.~\ref{fig1} for leviton's distribution function $f_{L}( \epsilon)$.   

Note that HL's energy is zero on average only but it does fluctuate.
This fact differs HL from quasi-particles in Majorana zero modes in topological insulators and superconductors, whose energy is strictly zero,  see, e.g., Refs.~\cite{Hasan:2010ku,Qi:2011hb}.
In addition HL is charged while a Majorana fermion is neutral. 

The distribution function for electron-hole pairs is an odd function of energy, $f_{eh}( \epsilon) = - f_{eh}(- \epsilon)$. 
Therefore, namely electron-hole pairs do carry (excess) energy, which is pumped by a time-dependent voltage $V(t)$ into the Fermi sea:

\begin{eqnarray}
\left\langle \epsilon \right\rangle_{eh} \equiv \int d \epsilon f_{eh}( \epsilon) \epsilon = i \hbar \int dt \frac{ \partial g_{eh}^{(1)}(t;t^{\prime}) }{ \partial t^{\prime} } \bigg|_{t^{\prime}=t} = \frac{1 }{4 } \frac{ \hbar }{ 2 \Gamma _{\tau} } .
\label{Qeh}
\end{eqnarray}
\ \\ \noindent 
This energy is a quarter of the energy of a leviton. 

The same result follows also from a time-dependent heat current, $J_{Q}(t)$, induced by a voltage pulse.\cite{Battista:2013ew} 
At zero temperature one can find quite generally \cite{Ludovico:2014de} that a charge current $I(t) = e g(t;t)$ and a heat current, both induced by a voltage pulse in a single-channel chiral or ballistic conductor, comply with the Joule law,

\begin{eqnarray}
J_{Q}(t) = R_{q} I^{2}(t)  ,
\label{JL}
\end{eqnarray}
\ \\ \noindent
where $R_{q} = h/(2 e^{2})$ is the charge relaxation resistance \cite{Buttiker:1993bw}, the B\"{u}ttiker resistance \cite{Glattli:2014tu}.  
Heat is nothing but the excess energy carried by excitations.\cite{Moskalets:2009dk} 
Indeed, $\int dt J_{Q}(t) = \hbar /(8  \Gamma _{\tau})$, which agrees with Eq.~(\ref{Qeh}).  

The fact that the Joule law, Eq.~(\ref{JL}), works in the present case is remarkable, since charge and heat are carried by different pieces of the excited state, HL and electron-hole pairs, respectively.  
Actually these pieces can be separated in experiment. 
To show this let us first consider what is excited by a voltage pulse of an opposite sign. 

{\it Anti-half-leviton.--}
In the case of $n^{\star}=-0.5$ the correlation function is  
$G_{-0.5} ^{(1)}(t_{1}; t_{2})  = e^{i \left( t_{1} - t_{2} \right) \frac{ \mu }{ \hbar } }   \left\{ -g_{HL}\left( t_{1}; t_{2} \right)  + g_{eh}^{(1)} \left( t_{1}; t_{2} \right) \right\} /  v_{ \mu} $. 
The change of a voltage sign does not alter an electron-hole part. 
What is changed is the sign of a charge of a single-electron excitation, which now I call {\it an anti-half-leviton} (aHL). 

Let us take the states with HL and aHL  excited at different contacts and mix them at a wave splitter, a quantum point contact, with transmission $T$ and reflection $R = 1 - T$ probabilities. 
The correlation function of excitations at output is 
$G_{out} ^{(1)} = T G_{0.5} ^{(1)} + R G_{-0.5} ^{(1)} $. 
In the case of a symmetric wave splitter, $T=R=0.5$, we find $G_{out} ^{(1)}(t_{1}; t_{2})  = e^{i \left( t_{1} - t_{2} \right) \frac{ \mu }{ \hbar } }  g_{eh}^{(1)} \left( t_{1}; t_{2} \right) /  v_{ \mu}$. 
That is, the state projected onto the output channel contains only an electron-hole state, see Fig.~\ref{fig2}.  
The measurement made on such a state can serve as the reference point for a  measurement made on $G_{0.5} ^{(1)}$ in order to extract characteristics of a half-leviton. 

Note that the excess first-order correlation function contains all information about excitations, their charge, energy, fluctuations, coherence times, etc. 
The correlation function is additive and, therefore, it is specifically suitable for the electron-hole pairs elimination procedure outlined above.  
The correlation function can be directly measured with the help of an interference current as it was suggested in Refs.~\onlinecite{Haack:2011em,Haack:2013ch}.  
Moreover, the distribution function $f_{HL}( \epsilon)$, Fig.~\ref{fig1}, can be measured using already available experimental tools, quantum dots as energy filters.\cite{Altimiras:2010ej}  
The level width of a quantum dot restricts the precision of measurement of a zero-energy peak.  
Another factor limiting a precision is a non-zero temperature of the Fermi sea. 

\begin{figure}[b]
\centerline{
\includegraphics[width=80mm]{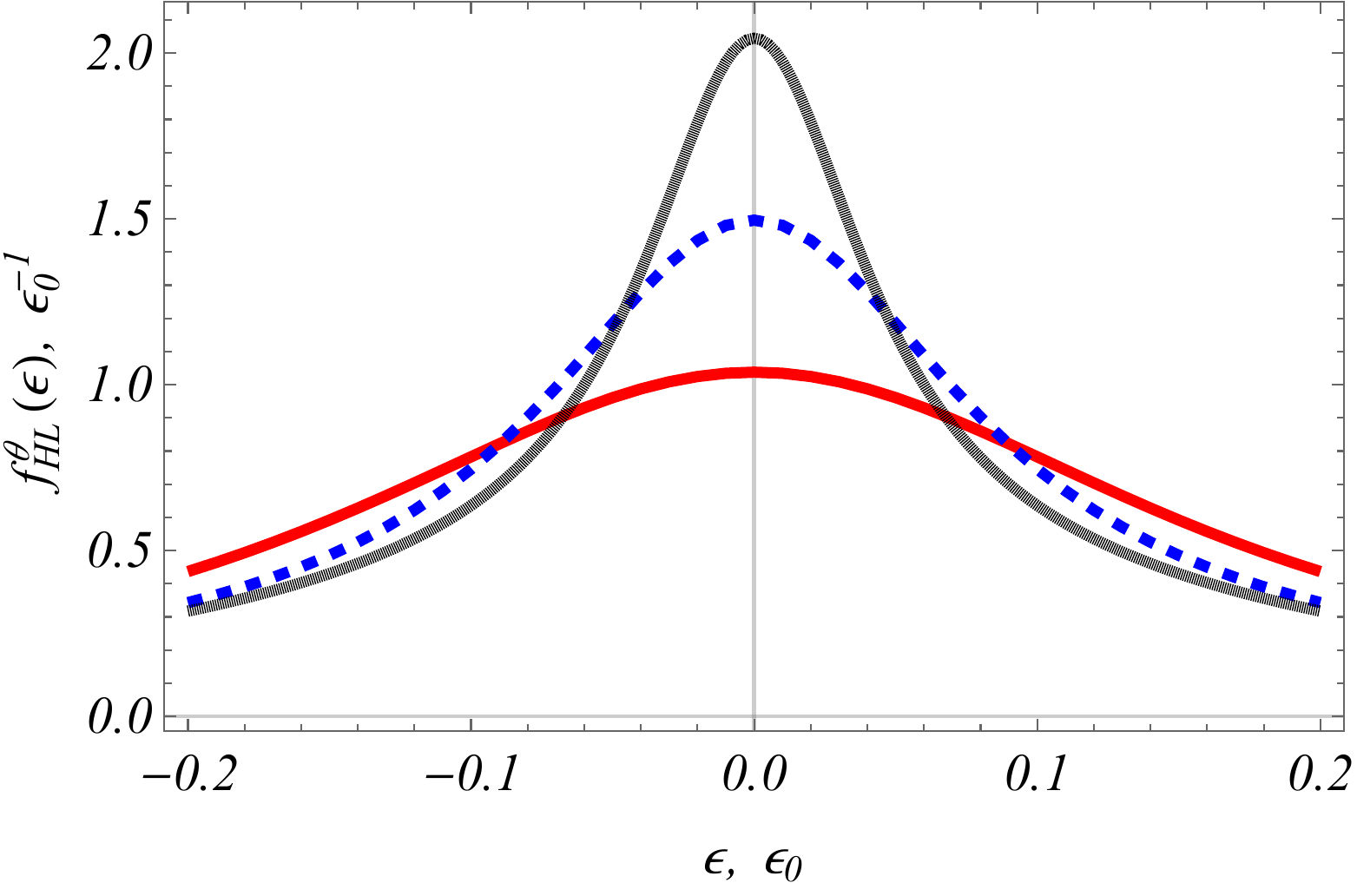}
}
\caption{(Color online) Energy distribution function $f_{HL}^{ \theta}( \epsilon)$, Eq.~(\ref{fHLth}), of a half-leviton at different temperatures of the Fermi sea $ \theta = \theta^{\star}/k$, where $\theta^{\star} = \hbar/( \pi k_{B}  \Gamma _{\tau})$ and $k=5$ (red solid line), $k=10$ (blue dashed line), and $k=20$ (black short-dashed line). 
For a voltage pulse with width $  2\Gamma _{\tau} = 30~{\rm ps}$ \cite{Dubois:2013ul} the characteristic temperature is $ \theta^{\star} \approx 160~{\rm mK}$.}
\label{fig3}
\end{figure}

{\it Temperature effect.--}
At a non-zero temperature, $ \theta > 0$, the excess correlation function  is,
$G ^{(1)}_{ 0.5, \theta}\left( t_{1};t_{2} \right) =  \eta\left( \frac{  t_{1} - t_{2} }{ \tau_{ \theta} }  \right)  G ^{(1)}_{ 0.5}\left( t_{1};t_{2} \right)$,
where $\eta(x) = x/\sinh(x)$ is a temperature-induced suppression factor and the thermal time $\tau_{ \theta} =\hbar /(  \pi k_{B} \theta )$  with $k_{B}$ the Boltzmann constant.\cite{Moskalets:2016dx}
Substituting the equation above into Eq.~(\ref{f}) and isolating a part related to HL we find,

\begin{eqnarray}
f_{HL}^{ \theta} ( \epsilon) = \int d \omega \eta_{ \omega}  f_{HL} ( \epsilon + \hbar \omega )  . 
\label{fHLth}
\end{eqnarray}
\ \\ \noindent
where $ \eta_{ \omega} = ( \pi \tau_{ \theta}/4) \cosh^{-2}( \pi \omega \tau_{ \theta} /2)$ is the Fourier transform of $ \eta(t/ \tau_{ \theta}) = \int d \omega e^{-i \omega t} \eta_{ \omega}$. 
A zero-temperature distribution function $f_{HL}( \epsilon)$ is given in Eq.~(\ref{fHL}). 
The function $f_{HL}^{ \theta} ( \epsilon)$ is presented in Fig.~\ref{fig3} for different temperatures of the Fermi sea. 
Though a zero-energy peak is suppressed with increasing temperature its shape remains symmetric around $ \epsilon=0$.   
 
{\it Conclusion.--}
Dynamically perturbed Fermi sea can host exotic zero-energy excitations with an effectively fractional charge. 
In this Letter I discussed an example of such a quasi-particle, which can be  excited by  a Lorentzian voltage pulse $V(t)$ with a half-integer Faraday flux, $\varphi = (e/ \hbar) \int dt V(t) = \pi$ using the same technique that was used to generate levitons \cite{Dubois:2013ul,Jullien:2014ii}. 
A single particle with an effective charge $e/2$, a half-leviton (HL), is excited together with a cloud of electron-hole pairs, which, however, can be isolated and used as the reference point for studying HL. 
A half-leviton is described by a single-particle state, which is mixed in equal proportions with the vacuum state hence a fractional charge. 
This single-particle state is a coherent superposition of states with energies symmetrically placed near the Fermi energy. 
Therefore, the energy of HL counted from the Fermi energy is zero. 
The wave function of HL is real-valued and, therefore, it remains the same when we go over to an anti-HL, a particle excited by a voltage pulse with $\varphi =  - \pi$. 
These properties enable HL and anti-HL to annihilate each other while colliding at an electronic wave splitter, what paves a way for entangling fermionic Fock states with different number of particles.   
Dynamic excitation of an electron many-particle system is an exciting and promising platform for quantum coherent electronics, which ``...does not require delicate nanolithography, considerably simplifying the circuitry for scalability'' \cite{Dubois:2013ul}.


\end{document}